\documentclass[twocolumn,english]{IEEEtran}
\usepackage[T1]{fontenc}
\usepackage[utf8]{inputenc}
\usepackage{listings}
\usepackage{verbatim}
\usepackage{varioref}
\usepackage{amsthm}
\usepackage{graphicx}

\makeatletter

\providecommand{\tabularnewline}{\\}

\theoremstyle{plain}
\newtheorem{thm}{\protect\theoremname}
\theoremstyle{definition}
\newtheorem{defn}[thm]{\protect\definitionname}

\usepackage{bbding}
\usepackage{multicol}
\usepackage{listings}

\makeatother

\usepackage{babel}
\providecommand{\definitionname}{Definition}
\providecommand{\theoremname}{Theorem}

\begin{document}

\title{Model Driven Mutation Applied to\\ Adaptative Systems Testing}

\author{Alexandre Bartel$^1$, Benoit Baudry$^2$, Freddy Munoz$^2$, Jacques Klein$^1$, Tejeddine Mouelhi$^1$ and Yves Le Traon$^1$\\ \vspace{.25cm} $^1$ Interdisciplinary Center for Security, Reliability and Trust\\ University of Luxembourg\\ L-1359 Luxembourg-Kirchberg, Luxembourg \\\{alexandre.bartel, jacques.klein, tejeddine.mouelhi, yves.letraon\}@uni.lu\\ \vspace{.25cm} $^2$ INRIA Centre Rennes - Bretagne Atlantique\\ Campus de Beaulieu\\ 35042 Rennes, France \\\{benoit.baudry, freddy.munoz\}@inria.fr}

\maketitle

\begin{abstract}
Dynamically Adaptive Systems modify their behavior and structure in
response to changes in their surrounding environment and according
to an adaptation logic. Critical systems increasingly incorporate
dynamic adaptation capabilities; examples include disaster relief
and space exploration systems. In this paper, we focus on mutation
testing of the adaptation logic. We propose a fault model for adaptation
logics that classifies faults into environmental completeness and
adaptation correctness. Since there are several adaptation logic languages
relying on the same underlying concepts, the fault model is expressed
independently from specific adaptation languages. Taking benefit from
model-driven engineering technology, we express these common concepts
in a metamodel and define the operational semantics of mutation operators
at this level. Mutation is applied on model elements and model transformations
are used to propagate these changes to a given adaptation policy in
the chosen formalism. Preliminary results on an adaptive web server
highlight the difficulty of killing mutants for adaptive systems,
and thus the difficulty of generating efficient tests. \end{abstract}
\thispagestyle{empty}

\section{Introduction}

Dynamically Adaptative Systems (DAS) must adapt themselves to ongoing
circumstances and find the way to continue accomplishing their functionalities.
DAS play increasingly important role in society’s infrastructures;
the demand for DAS appears in application domains ranging from crisis
management applications such as disaster management \cite{Hughes_anintelligent},
space exploration \cite{Dvo99}, and transportation control to entertainment
and business applications. This demand is intensified by the mobile
and nomadic nature of many of these domains. The IDC%
\footnote{IDC is an analyst company and a global provider of market intelligence,
advisory services, and events for the information technology, telecommunications,
and consumer technology markets.%
} analysts forecast a global increase in the number of mobile workers
to more than 1.19 billion by 2013 \cite{Drake2010}.

DAS respond to environmental changes by modifying their internal configuration
to continue meeting their functional and non-functional requirements.
Designing a DAS involves specifying environmental fluctuations that
have an impact on the system, as well as the related strategies for
performing the structural changes. This is captured by an adaptation
logic that expresses the actions to be adopted when the environment
changes \cite{chauvel:2008c,Eliassen.2006.2,keeney03chisel,oai:CiteSeerXPSU:10.1.1.94.6543}.
More precisely, adaptation logics drive the adaptation process and
compute the right system configuration that should be adopted given
an environmental condition.

This paper focuses on the issue of testing whether an adaptation logic
is correctly implemented. More specifically, we focus on mutation
of adaptation logic, considering that test cases should be able to
distinguish between the original adaptation logic and the mutated
one. Mutation thus provides a qualification criterion for test cases.

We use a Model-driven engineering (MDE) process to model adaptation
formalisms/languages as well as adaptation policies defined according
to these formalisms. A metamodel captures all the necessary concepts
for representing action-based adaptation policies. From the metamodel,
we derive mutation operators that can apply to several action-based
adaptation formalisms. 

We classify adaptation logic faults into two groups:
\begin{enumerate}
\item The possible environmental conditions the system will face, and
\item the complexity involved in producing a response to those conditions.
\end{enumerate}
The first, environmental completeness (EC) faults embody faults due
to gaps in the space covered by the adaptation logic, thereby missing
adaptations for environmental changes. The second, adaptation correctness
(AC) faults embody faults due to incorrect adaptations to environmental
changes. Our hypothesis is that managing environmental changes involving
a single property variation (simple) is easy, whereas managing several
properties varying at the same time (complex) is error prone. We summarize
the contributions of this paper as follows:
\begin{enumerate}
\item A generic metamodel capturing the concepts inherent to adaptation
logic, completed with model transformations from two different input
formalisms.
\item A generic set of mutation operators for adaptation logics as well
as a specialization of this model to action-based adaptation logics.
\item A first proof of concepts through an adaptive web server case study.
\end{enumerate}
It has to be noted that we do not deal with efficient test cases generation
in this paper, and for the experiments we simply create test sequences
randomly (sequence of events issued by the environment).

The remainder of this paper proceeds as follows. Section 2 provides
a background on dynamically adaptive systems. Section 3 introduces
model driven engineering techniques and explains how they can be used
with testing adaptation logics. Section 4 describes the first mutation
operators we used. Section 5 presents our first experiments. Section
6 presents the related work. Finally, we conclude and present our
perspectives in section 7.

\section{Dynamically adaptive systems }

Consider an adaptive web server, which processes file requests over
the HTTP protocol. It answers these requests as fast as possible while
optimizing the resources it consumes, e.g. memory, CPU time, etc.
Additionally, it provides non-stop service, thus it needs to adapt
its internal structure to respond to a changing working environment.
This environment is characterized by the variable amount of requests
over time.

\subsection{Environment and configurations }

Dynamically adaptive systems (DAS) encode the environment into an
abstraction called context. 
\begin{defn}[context]
\label{def:-A-context} A context consists of a n-tuple of fields
<$p_0$,$p_1$, …, $p_n$>, where each field $p_i$ represents an
environmental property. The type of each field is defined by the encoding
chosen for the property it represents. 
\end{defn}
In our adaptive web server example, the environment is modeled as
a context with two properties:
\begin{itemize}
\item $p_1$: number of request per second (server load); 
\item $p_2$: the percentage of request (request density). 
\end{itemize}
The last one corresponds to the number of requested files. The domain
or type of each property has a lower and an upper bound. For instance,
we associate the type integer with request density and server load,
a lower bound 0 and upper bound 100 for both. The server load domain
indicates that the minimum amount of request in one second is 0 (no
request) and the maximal is 100. Analogous, request density indicates
the number of requested files.

\begin{defn}
Specific environmental conditions at an instant t are drawn by an
instance I of the context representing the environment. Such an instance
is an n-tuple of values corresponding to the punctual value of a particular
property. 
\end{defn}
The context instance <12, 3> designates a particular environmental
condition with 12 requests per second requesting 3 different files.
A sequence of context instances $I_0$, $I_1$, $I_2$, …, $I_n$
ordered by their occurrence over time is called a context flow (CF).
A context generates a space containing all the possible instances
that can produce the combination of property values. The context of
the adaptive web server generates a space containing all its possible
context instances.

\subsection{Adaptation logic }

Adaptation in DAS is driven by an adaptation logic (adaptation model)
that uses a specific strategy to describe the configuration to adopt
given a context change. 
\begin{defn}
\label{def:An-adaptation-logic}An adaptation logic defines a relation
between contexts and system configuration. It receives a context
instance (current environmental condition), a context flow (history
of the environment, and the history of the system configuration, and
gives the next configuration the system must adopt.
\end{defn}
There exist several strategies to describe adaptation logics, a few
examples are: action-based adaptation \cite{keeney03chisel}, where
adaptations are triggered when a condition is satisfied; goal based
adaptation \cite{Eliassen.2006.2}, where adaptations are performed
to reach a specific goal; and utility function based adaptation \cite{oai:CiteSeerXPSU:10.1.1.94.6543},
where adaptations are calculated according to a cost function based
on environmental conditions and variation point value.

An action-based strategy describes the adaptation logic of the adaptive
web server \cite{keeney03chisel}. In this case the adaptation logic
is a set of rules (adaptation policies) that, whenever an event occurs
(environmental change) evaluate if a set of conditions are satisfied,
and if it is so, they perform a series of adaptation actions. 

\begin{table}[tbh]
\caption{\label{tab:Excerpt-of-the}Excerpt of the adaptive web server adaptation
logic.}

\inputencoding{latin9}\begin{lstlisting}
1: when requestdensity is 'high' or 'medium'
2: if cacheHandler.size  == 0 
3: then utility of addCache is 'high'    

5: when requestdensity is 'low'  
6: if cacheHandler.size == 0  
7: then utility of addCache is 'low'

9: when LOAD is 'high'
10: if FileServers.size  <= 10 
11: then utility of addFileServer is 'high'  

12: when LOAD is 'LOW'  
13: if FileServers.size  <= 10 
14: then utility of addFileServer is 'low'
\end{lstlisting}
\inputencoding{utf8}

\end{table}

Table \vref{tab:Excerpt-of-the} presents an excerpt of the adaptive
web server adaptation logic. The first two rules manage the system
cache.  The first rule (lines 1-3) enables the cache mechanism when
the request density is high or medium (line 1) and there is no cache
(line 2).  The second rule (lines 5-6) is analogous to the first.
It reflects the fact that when the dispersion is high, adding a cache
is not very useful. The remaining rules (lines 9-14) handle the variations
of the server load property.

While table 1 presents a textual action-based adaptation logic, the
Diva framework%
\footnote{http://www.ict-diva.eu/DiVA/%
} allows to express the adaption policy in the form of a set of tables
which are directly manipulated as models elements. Thus the connection
with our MDE process is natural.

\section{MDE and adaptation logics}

This section will introduce the MDE concepts which are required to
understand how we create and use mutation operators later in the paper.

\subsection{Metamodeling, Kermeta and Sintaks}

This section summarizes the intents of metamodelling and how the \textit{Kermeta}
environment fits in this modelling activity.

\subsubsection{Metamodelling}

Metamodelling is a technique used to build a metamodel that defines
a modeling language for a particular domain. The metamodel defines
the concepts and relationships that describe the domain. A metamodel
is itself a model expressed with a modeling language called the meta-metamodel.

\subsubsection{Kermeta }

Kermeta \cite{Mull05a} is a metamodelling environment developed at
IRISA (Research Institute in Information technology and rAndom Systems).
This imperative and object-oriented language is used to provide an
implementation of operations defined in metamodels.

\subsubsection{Sintaks}

Sintaks \cite{Muller06a} is a tool to defines bridges between plain
text files and models.

\subsection{Action-based adaptation logic metamodel}

\begin{figure*}[tbh]
\begin{centering}
\includegraphics[width=2\columnwidth,height=1.3\columnwidth]{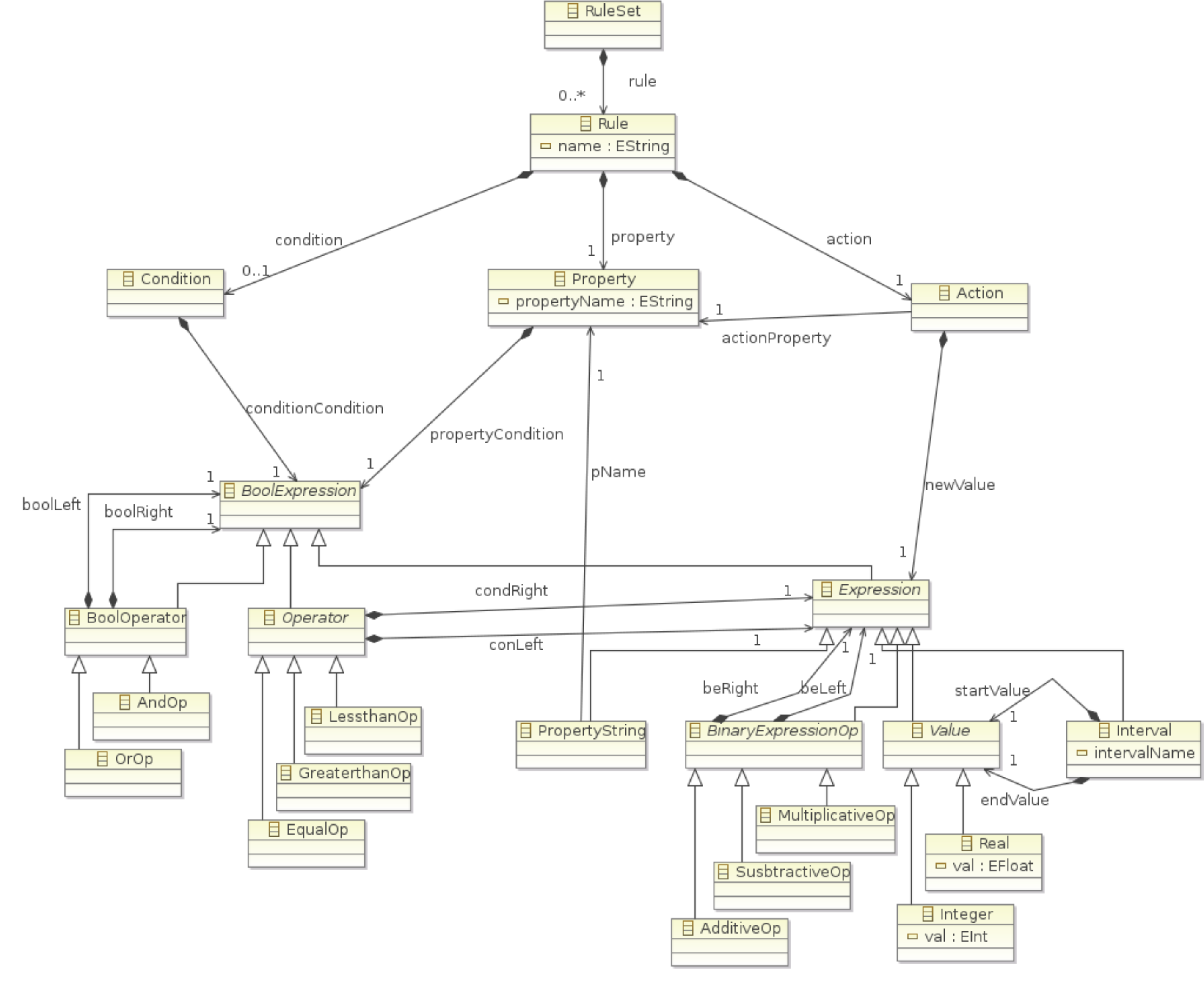}
\par\end{centering}

\caption{\label{fig:Metamodel-of-the}Metamodel of action-based adaptation
logics}
\end{figure*}

Figure \ref{fig:Metamodel-of-the} represents the metamodel we propose
to represent the abstraction of action-based adaptation logics. An
action-based adaptation logic always consists of a set of rules (RuleSet
in the figure) called Event-Condition-Action, or ECA, rules. One ECA
rule (Rule) features one event (Property), one condition (Condition)
and one action (Action). An event is bound to a context property.
When the bounded context property changes and its new value matches
the event condition (propertyCondition), then the rule is executed.
When the rule is executed the condition (Condition->BoolExpression)
has to be true to perform the rule's action. This condition usually
refers to internal states of the adaptation system. The action consists
of assigning a new value (newValue) to a property (actionProperty). 

In short, a rule performs an Action if the bounded Event property
in the new context matches the Event condition and if the rule Condition
is true. For instance, the first rule of the adaptation logic represented
table \ref{tab:Excerpt-of-the}, is bounded to the property {}``requestdensity''.
The rule will be executed only after specific context changes in which
the property became {}``high'' or {}``medium''. The action of
assigning {}``high'' to {}``addCache'' is only performed if the
internal variable cacheHandle.size equals zero.

The metamodel ecore file was created using EMF (Eclipse Modeling Framework)
and GEE (Graphical Ecore Editor). 

We will describe the process of mutant generation as well as the genericity
of the metamodel in the following section.

\subsection{Generic process for mutant generation}

\begin{figure}[tbh]
\begin{centering}
\includegraphics[width=0.9\columnwidth]{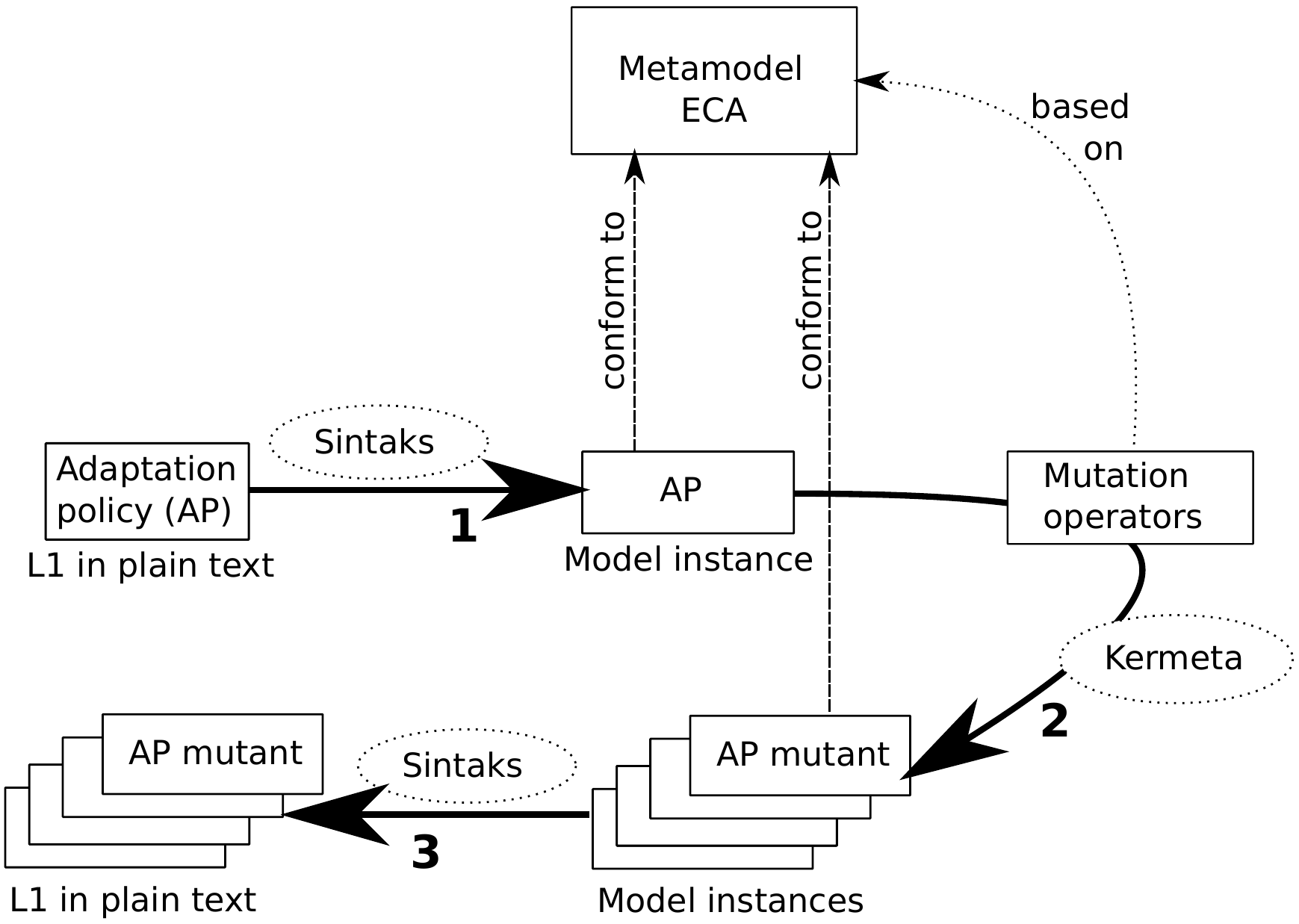}
\par\end{centering}

\caption{\label{fig:Mutant-generation-process.}Mutants generation process.}

\end{figure}

Figure \vref{fig:Mutant-generation-process.} represents the mutants
creation process. The process start by selecting an adaptation policy
(AP) expressed in an action-based language (L1 on the figure). The
first step (1) is to transform the adaptation policy into a model
conforming to the metamodel. The second step consists in applying
the mutation operators to the policy. Mutation operators are generic
and work on models, not plain text files. Once the mutant models have
been generated, they are transformed in plain text files. The Sintaks
tool was used to do a mapping between a rule set written in plain
text and its model representation.

Since mutation operators are defined from the metamodel and are working
on models, they are independent of the action-based language used
to write the adaptation policy: a bridge between textual files and
models must be defined for each language. This is achieved by defining
one bridge for each language with Sintaks.

As a result we got a set ofResulting plaintext mutants will be used
to test the adaptation logic's test suites. We consider that test
suites must be able to distinguish a correct adaptation logic from
the incorrect ones. 

In the following section we introduce the first mutation operators.

\section{Mutation operators for adaptation logics}

Definition \ref{def:An-adaptation-logic} introduces the concept of
adaptation logic as the driver of the adaptation. Testing the realization
of such driver means verifying whether the system is capable of adapting
to environmental changes, and whether such adaptations proceed as
expected. This section presents the challenges associated with testing
adaptation logics, as well as a fault model for adaptation logics.

\subsection{Testing challenges }

Testing adaptation logics involves generating context instances, and
evaluating the results of exposing the system to such context instances. 

Three steps compose the testing process: 
\begin{enumerate}
\item Initially, testers synthesize a context flow from a series of context
instances. 
\item Then, they execute and expose the system to the generated context
flow. Testers evaluate whether the  configurations adopted by the
system (configuration flow) when exposed to environmental changes
are as expected. If not, the adaptation logic contains a fault. 
\item The process may start again until a qualification criterion is reached.

\end{enumerate}
Note that (1) and (2) are not the object of the paper. Thus we generate
test cases randomly. We rather focus on (3).

A \textit{test suite} is a set of\textit{ test cases}. In this paper
a test case is defined as a \textit{context flow} of a certain length,
L. L represents the number of \textit{context instances} in the flow.
Given a flow f containing L context instances $I_i$ $i \in (1,2,...,L)$,
$I_i$ and $I_{i+1}$ differs by one or more of their properties'
values. For each $I_i$ the adaptive system will generate one ore
more events corresponding to the properties that have changed. Those
events are then handled by the adaptation logic (rules) which generates
a new configuration for the system.

A test case is said to kill a mutant if the result (new configuration)
generated by the mutant adaptation logic differs from the result given
by the original adaptation logic.

This process enables us to detect:
\begin{itemize}
\item duplicate rules or useless rules (the mutant is not killed in this
case)
\item errors in the adaptation logic

\begin{itemize}
\item either an event in not handled properly or
\item an incorrect action in performed leading to an incorrect new configuration
\end{itemize}
\end{itemize}

\subsection{Fault model for adaptation logics }

Managing the scenarios to which a system adapts is complex due to
their large number and the difficulty to foresee the interactions
between them. 

In this section we introduce  generic mutation operators for the
adaptation logics metamodel.   Those operators will mutate adaptation
logic models conforming to the metamodel and thus are independent
of any adaptation logic language.

\subsubsection{Environmental completeness faults }

Definition \ref{def:-A-context} defines a context as a tuple of fields
representing environmental properties. The adaptation logic interprets
these fields’ values, and decides the system configuration that best
fits the environmental conditions. It is possible, however, that the
adaptation logic neglects some property values, or a complete property.
We call faults of this type environmental completeness (EC) faults.

In the following, we describe three different types of EC faults represented
as mutation operators.

\begin{enumerate}
\item ICP - Ignore Context Property

\vspace{.25cm}

\fbox{\begin{minipage}{0.86\columnwidth}
For a given property p, delete each rule that can be executed on p.
\end{minipage}}

\vspace{.25cm}

For instance, when ignoring property {}``requestdensity'' the two
last rules in table \ref{tab:Excerpt-of-the} (lines 9-14) are deleted.

\item ISV - Ignore Specific Context Property Value

\vspace{.25cm}

\fbox{\begin{minipage}{0.86\columnwidth}
For a given couple (property p, value v), delete each rule that can be executed when p equals v.
\end{minipage}}

\vspace{.25cm}

When ignoring value {}``high'' for property {}``LOAD'' one rule
(lines 9-11) is deleted.

\item IMV - Ignore Multiple Context Property Values

\vspace{.25cm}

\fbox{\begin{minipage}{0.86\columnwidth}
For a given set of couples (property $p_i$, value $v_i$), delete each rule that can be executed when any $p_i$ (i $\in$ \{1,2,...,N\}) equals $v_i$. (At least two rules with different properties are modified/deleted).
\end{minipage}}

\vspace{.25cm}

When ignoring value {}``high'' for property {}``LOAD'' and {}``low''
for property {}``requestdensity'', two rules (lines 5-7 and lines
9-11) are deleted.

\end{enumerate}

\subsubsection{Adaptation correctness faults }

The observable behavior produced by the adaptation logic is the adaptation
it produces facing an environmental change. Some times such adaptation
does not change the system in the expected way. We call this kind
of faults adaptation corrected (AC) faults, because they lead directly
to incorrect adaptations. Notice that the observable behavior of EC
faults is manifested in at least one of the following AC faults.

\begin{enumerate}
\item SRA - Swap Rule Action

\vspace{.25cm}

\fbox{\begin{minipage}{0.86\columnwidth}
The action values from two rules modifying the same property are swapped.
\end{minipage}}

\vspace{.25cm}

For instance {}``high'' and {}``low'' swap lines 11 and 14 in
table \ref{tab:Excerpt-of-the} for property {}``addFileServer''.

\item Modify Rule Condition Value

\vspace{.25cm}

\fbox{\begin{minipage}{0.86\columnwidth}
The condition value (always on the right part of the condition), for a condition which uses operator > or <, in a rule is decreased or increased, respectively.
\end{minipage}}

\vspace{.25cm}

For instance in table \ref{tab:Excerpt-of-the} line 10, the value
{}``10'' is increased to {}``100''.

\end{enumerate}

\section{Experiments }

In this section we present a preliminary proof of concept based on
the adaptative web server system.

\subsection{Test subject }

To validate our hypothesis about the ability of AST to uncover faults
in adaptation logics, we use the adaptive web server presented in
section 2 as a test subject. 

\begin{figure}[tbh]
\begin{centering}
\includegraphics{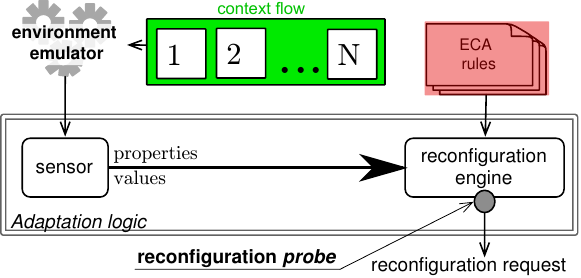}
\par\end{centering}

\caption{\label{fig:Instrumented-architecture-of}Instrumented architecture
of the adaptive web server adaptation logic}

\end{figure}

Figure \ref{fig:Instrumented-architecture-of} illustrates the architectural
realization of the adaptation logic presented in section 2. It is
composed of a sensor component, which is aware of the environment
and collects the data produced by environmental changes. It encodes
the data into values representing the environmental properties of
interest (context instance) and passes them to a  reconfiguration
engine. Finally, the reconfiguration engine loads the adaptation rules
and matches the values against the adaptation rules. If an adaptation
rule matches the values, then it requests the system implementation
to reconfigure as described by the rule. 

To inject context instances and collect reconfiguration data we have
instrumented the adaptation logic. Figure \vref{fig:Instrumented-architecture-of}
presents the instrumented architecture. We have modified the source
code of the sensor component and replaced the environment sensing
mechanism with an environment emulator. This emulator reads context
flows from a text file and injects them into the system provoking
the instrumented sensor to respond identically to the non-instrumented
one. We have also added a reconfiguration probe that records the reconfiguration
requests produced by the reconfiguration engine.

\subsection{Experiment set up, results and analysis}

\subsubsection{Experiment}

\begin{table}[tbh]
\caption{\label{tab:Experiment-set-up-and}Experiment set-up and execution}

\begin{centering}
\begin{tabular}{|l|c|}
\hline 
\# of test suites & 30\tabularnewline
\hline 
\# of context flows per test suite & 10\tabularnewline
\hline 
\# of context instances per flow & 20\tabularnewline
\hline 
\# of mutants of the adaptation logic & 130\tabularnewline
\hline 
\hline 
Total number of simulations & 39.000 ($ 30 \cdot 10 \cdot 20 \cdot 130$)\tabularnewline
\hline 
\end{tabular}
\par\end{centering}

\end{table}

We prepared and executed our experiment as described in table \vref{tab:Experiment-set-up-and}.
We generated 30 test suites. Each of them contains 10 test cases (context
flows). A flow is created by uniformly selecting a sequence of context
instances among all the possible context instances.

\subsubsection{Results and analysis}

\begin{table}[tbh]
\caption{\label{tab:Experiment-results}Experiment results}

\centering{}%
\begin{tabular}{|l|c|}
\hline 
\multicolumn{1}{|l|}{Test suite} & Random\tabularnewline
\hline 
\hline 
minimun mutation score & 91/130 $\approx$ 70\%\tabularnewline
\hline 
maximum mutation score & 96/130 $\approx$ 74\%\tabularnewline
\hline 
average mutation score & 93/130 $\approx$ 71\%\tabularnewline
\hline 
\end{tabular}
\end{table}

Table \vref{tab:Experiment-results} presents the global mutation
score (number of unique killed mutants). 

What we notice is that 30\% of the mutants are not killed with random-generation.
Even if we take longer test cases the results are similar. This first
shows that other techniques should be studied.

\subsection{Threats to validity }

There exists no perfect data, or perfectly trustable analysis results,
and this study is not an exception. For this reason we identify the
construction, internal and external threats to validity for this study.

Internal threats lie on the source and nature of the empirical data.
We recognize that we have only studied a small adaptive system realizing
the adaptation logic through action-based reasoning. The limited number
of environmental properties, and the size of the space represent a
threat since it is easy to achieve a uniform coverage with few context
instances. 

External threats lie on the statistical significance of our study.
We are aware that since the adaptive system is small and only one,
it does not represent the industrial trends. To make more general
statements it is necessary to try the presented technique on large
system. However, DAS are an emergent technology still paving its adoption.

\section{Related work }

As far as we know there is no other work that uses mutation to measure
the quality of adaptation logics' tests. However, a large number of
researchers have addressed the validation and testing problem of adaptive
systems. Zhang et al. \cite{conf/aosd/ZhangGC09} address the verification
of dynamically adaptive systems through modular model checking. They
model the adaptive system as finite state machine in which states
represent different system variants. Zhang and Cheng \cite{conf/icse/ZhangC06}
introduce a model-based development process for adaptive software
that uses Petri-nets. Biyani and Kulkarni \cite{conf/icse/BiyaniK07}
use predicate detection for testing adaptive systems during adaptation.
They extend existing algorithms based of global predicate evaluation
\cite{BabaogluOz1993a} for testing distributed systems to the system
during adaptation. Kulkarni and Biyani\cite{conf/cbse/KulkarniB04}
introduce an approach using proof-lattice to verify that all possible
adaptation paths do not violate global constraints. Allen et al \cite{1998:fase:allen}
used the Wright ADL to integrate the specifications of both architectural
and behavioral aspects of dynamically reconfigurable systems. Kramer
and Magee \cite{journals/iee/KramerM98} use property automata and
labeled transition systems to specify and verify adaptive program’s
properties. The main difference between these verification approaches
and ours is the focus of attention. We are interested in verifying
through testing the adaptation driver, and not the adaptation process
itself. Furthermore, these approaches require computing the entire
system configurations and the transitions between them, however sometimes
this is not possible. 

Lu et al. \cite{conf/icse/LuCT08} study the testing of pervasive
context-aware software. They propose a family of test adequacy criteria
that measure the quality of test sets with respect to the context
variability.

Since very different testing techniques exist, we hope that mutation
will reveal itself as a good way to compare them.

\section{Conclusions and Perspectives}

The mutation operators presented in this paper are a first proposal
to offer a qualification environment for comparing testing techniques
applied to action-based adaptative systems. The use of MDE makes it
possible to derive mutants for most action-based logics, thus providing
a common framework for such test cases qualification. The case study
shows the feasibility of the approach and confirms that, for killing
mutants, other testing techniques should be considered rather than
random test generation. Due to the size of the case study and the
number of environmental properties it contains, it is not possible
to generalize to larger DAS. Future work will %
\begin{comment}
what is our notion of completeness? need to formalize.
\end{comment}
thus consist of completing the set of mutation operators, and will
exhibit experimental results on other case studies comparing several
test generation techniques. We plan to experiment with a much larger
case study, which comprises several environmental properties and interactions.
Furthermore we plan studying and specializing our fault model to other
adaptation logic technologies, such as goal oriented.

\bibliographystyle{abbrv}%plain}
\bibliography{bib/biblio}

\begin{thebibliography}{10}

\bibitem{1998:fase:allen}
R.~Allen, R.~Douence, and D.~Garlan.
\newblock Specifying and analyzing dynamic software architectures.
\newblock pages 21--37.

\bibitem{BabaogluOz1993a}
O.~Babaoglu and K.~Marzullo.
\newblock Consistent global states of distributed systems: Fundamental concepts
  and mechanisms.
\newblock Technical Report UBLCS-93-1, University of Bologna, Department of
  Computer Science, Jan. 1993.

\bibitem{conf/icse/BiyaniK07}
K.~N. Biyani and S.~S. Kulkarni.
\newblock Testing dynamic adaptation in distributed systems.
\newblock In H.~Zhu, W.~E. Wong, and A.~M. Paradkar, editors, {\em AST}, pages
  51--54. IEEE, 2007.

\bibitem{chauvel:2008c}
F.~Chauvel, O.~Barais, I.~Borne, and J.-M. Jézéquel.
\newblock Composition of qualitative adaptation policies.
\newblock In {\em Automated Software Engineering Conference (ASE 2008)}, pages
  455--458, 2008.
\newblock Short paper.

\bibitem{Drake2010}
S.~D. Drake, R.~Boggs, and J.~Jaffe.
\newblock Worldwide mobile worken population 2009-2013 forecast, 2010.

\bibitem{Dvo99}
D.~Dvorak, R.~Rasmussen, G.~Reeves, and A.~Sacks.
\newblock Software architecture themes in {JPL}'s mission data system.
\newblock In {\em AIAA Space Technology Conference and Exposition, Albuquerque,
  NM.}, 1999.

\bibitem{Eliassen.2006.2}
F.~Eliassen, E.~Gj{\o}rven, V.~S.~W. Eide, and J.~A. Michaelsen.
\newblock Evolving self-adaptive services using planning-based reflective
  middleware.
\newblock In N.~V. Geoff~Coulson, editor, {\em The 5th annual Workshop on
  Adaptive and Reflective Middleware (ARM 2006)}, pages 1--6. ACM Press, 2006.

\bibitem{Hughes_anintelligent}
D.~Hughes, P.~Greenwood, G.~Blair, G.~Coulson, P.~Smith, and K.~Beven.
\newblock An intelligent and adaptable grid-based flood monitoring and warning
  system.
\newblock In {\em Proceedings of the UK eScience All Hands Meeting}, pages
  53--60, 2005.

\bibitem{keeney03chisel}
J.~Keeney and V.~Cahill.
\newblock Chisel: {A} policy-driven, context-aware, dynamic adaptation
  framework.
\newblock In {\em Proceedings of the $4^{th}$ IEEE International Workshop on
  Policies for Distributed Systems and Networks (Policy 2003)}, pages 3--14.
  IEEE, June 2003.

\bibitem{journals/iee/KramerM98}
J.~Kramer and J.~Magee.
\newblock Analysing dynamic change in distributed software architectures.
\newblock {\em IEE Proceedings - Software}, 145(5):146--154, 1998.

\bibitem{conf/cbse/KulkarniB04}
S.~S. Kulkarni and K.~N. Biyani.
\newblock Correctness of component-based adaptation.
\newblock In I.~Crnkovic, J.~A. Stafford, H.~W. Schmidt, and K.~C. Wallnau,
  editors, {\em CBSE}, volume 3054 of {\em Lecture Notes in Computer Science},
  pages 48--58. Springer, 2004.

\bibitem{conf/icse/LuCT08}
H.~Lu, W.~K. Chan, and T.~H. Tse.
\newblock Testing pervasive software in the presence of context inconsistency
  resolution services.
\newblock In W.~Sch{\"a}fer, M.~B. Dwyer, and V.~Gruhn, editors, {\em 30th
  International Conference on Software Engineering ({ICSE} 2008), Leipzig,
  Germany, May 10-18, 2008}, pages 61--70. ACM, 2008.

\bibitem{Muller06a}
P.-A. Muller, F.~Fleurey, F.~Fondement, M.~Hassenforder, R.~Schneckenburger,
  S.~Gérard, and J.-M. Jézéquel.
\newblock Model-driven analysis and synthesis of concrete syntax.
\newblock In {\em Proceedings of the MoDELS/UML 2006}, Genova, Italy, Oct.
  2006.

\bibitem{Mull05a}
P.-A. Muller, F.~Fleurey, and J.-M. J\'ez\'equel.
\newblock Weaving executability into object-oriented meta-languages.
\newblock In S.~K.~L. Briand, editor, {\em Proceedings of {MODELS/UML}'2005},
  volume 3713 of {\em LNCS}, pages 264--278, Montego Bay, Jamaica, Oct. 2005.
  Springer.

\bibitem{oai:CiteSeerXPSU:10.1.1.94.6543}
W.~E. Walsh, G.~Tesauro, J.~O. Kephart, and R.~Das.
\newblock Utility functions in autonomic systems, 2004.

\bibitem{conf/icse/ZhangC06}
J.~Zhang and B.~H.~C. Cheng.
\newblock Model-based development of dynamically adaptive software.
\newblock In L.~J. Osterweil, H.~D. Rombach, and M.~L. Soffa, editors, {\em
  ICSE}, pages 371--380. ACM, 2006.

\bibitem{conf/aosd/ZhangGC09}
J.~Zhang, H.~Goldsby, and B.~H.~C. Cheng.
\newblock Modular verification of dynamically adaptive systems.
\newblock In {\em Proceedings of the 8th International Conference on
  Aspect-Oriented Software Development, {AOSD} 2009, Charlottesville, Virginia,
  {USA}, March 2-6, 2009}, pages 161--172. ACM, 2009.

\end{thebibliography}

\end{document}